\documentclass[letter,twocolumn]{jpsj2} %% two-column layout
\usepackage{bm} %bold math

\title{Antiferromagnetism of SrFe$_2$As$_2$ studied by
Single-Crystal $^{75}$As-NMR}

\author{Kentaro~\textsc{Kitagawa$^1$}\thanks{kitag@issp.u-tokyo.ac.jp},
Naoyuki~\textsc{Katayama$^1$}\thanks{present address: Department of Physics,
University of Virginia, Charlottesville, Virginia, USA},
Kenya~\textsc{Ohgushi$^{1,2}$}, and Masashi~\textsc{Takigawa$^{1,2}$}}

\inst{$^1$Institute for Solid State Physics, University of Tokyo,
5-1-5 Kashiwanoha, Kashiwa, Chiba 277-8581\\
$^2$JST, TRIP, 5 Sanbancho, Chiyoda, Tokyo 102-0075}

\abst{We report results of $^{75}$As nuclear magnetic resonance (NMR) 
experiments on a self-flux grown high-quality single crystal of SrFe$_2$As$_2$.
The NMR spectra clearly show sharp first-order antiferromagnetic (AF) and
structural transitions occurring simultaneously.
The behavior in the vicinity of the transition is compared with our previous
study on BaFe$_2$As$_2$. No
significant difference was observed in the temperature dependence of the static quantities such as the AF splitting and
electric quadrupole splitting. However, the results of the NMR relaxation rate
revealed difference in the dynamical spin fluctuations.
The stripe-type AF fluctuations in the paramagnetic state appear to be more
anisotropic in BaFe$_2$As$_2$ than in SrFe$_2$As$_2$.}
\kword{SrFe$_2$As$_2$, BaFe$_2$As$_2$, NMR, itinerant antiferromagnetism, ternary iron arsenide}

\begin{document}
\maketitle

%%\section{Introduction} 
The relevance of antiferromagnetism to the pairing mechanism of the
high-temperature cuprate superconductivity has been intensively discussed for
decades.
Last year, a new series of high temperature superconductors containing
iron pnictide layers has been discovered\cite{KamiharaJACS}.
Among them, the ternary compounds with ThCr$_2$Si$_2$
structure, $A_{1-x}B_x$Fe$_2$As$_2$ ($A$=Ba, Sr, Ca; $B$=K,
Na)\cite{RotterBa122SDW,RotterBK122,SasmalSrK122,WuCaNa122},
have particularly important features:
large crystals can be grown by flux methods and the undoped parent compounds
$A$Fe$_2$As$_2$ are reported to become superconducting under high
pressure\cite{KotegawaSr122HP,Alireza122HP, YuCa122HeHP,FukazawaBa122HP}.
Thus these materials provide an opportunity to 
investigate evolution from the antiferromagnetic (AF) to the superconducting
states without introducing disorder.

At ambient pressure $A$Fe$_2$As$_2$ shows structural and AF transitions
at the same temperature ($T_\text{t} \simeq 200$~K
for $A=$Sr and $\simeq 135$~K for $A=$Ba\cite{RotterBa122SDW}).
The structure is tetragonal with the space group $I4/mmm$ at room
temperature, and turns to the orthorhombic $Fmmm$ structure below $T_\text{t}$.
The low-temperature phase has a commensurate stripe-type AF
order\cite{HuangBa122NS,KanekoSr122NS}.
Because superconductivity appears when the AF transition is suppressed by
substituting Sr/Ba with K or by applying pressure, it is important to clarify
the nature of AF fluctuations in the parent compounds.
Previously, we have performed $^{75}$As nuclear magnetic
resonance (NMR) experiments on a self-flux grown single crystal of
BaFe$_2$As$_2$\cite{KitagawaBa122}.
The results of the spin-lattice relaxation rate $T_1^{-1}$ indicate
development of anisotropic spin fluctuations of stripe-type in the paramagnetic state.
In this letter, we report the $^{75}$As-NMR experiments on
a self-flux grown high-quality single crystal of SrFe$_2$As$_2$, which is
another member of the ternary series.
%We improve the method of the analysis on the
%relaxation rates specific to the $^{75}$As-NMR in $A$Fe$_2$As$_2$ lattice.
%The most important point is that the stripe AF fluctuations act on the
%relaxation rate only via the off-diagonal term in the hyperfine tensor.
We discuss different behavior of AF fluctuations between BaFe$_2$As$_2$ and
SrFe$_2$As$_2$.

%%\section{Experiment}
The single crystals of SrFe$_2$As$_2$ were prepared by
the self-flux method. The starting elements were mixed in a
alumina crucible with the ratio Sr:Fe:As=1:5:5 and sealed in a double quartz
tube. Excess FeAs works as flux. 
We put Zr sponge as a getter in the outer tube and sealed with Ar
gas in order to avoid contamination by air diffusing through the quartz wall.
The tube was heated up to 1100{$^\circ$}C in 14 hours (including the holding at 700{$^\circ$}C for 3 hours)
and slowly cooled down to 900{$^\circ$}C in 50 hours.
The resistivity and the magnetic susceptibility showed a sharp transition at
199~K, in agreement with the results by
Yan~\textit{et\,al.}\cite{YanSr122SC}

For NMR experiments, a crystal with the size $3\times 2\times
0.15$~mm$^{3}$ was mounted on a two-axis goniometer, which allows fine 
alignment of crystalline axes along the magnetic field within
0.2{$^\circ$}. The field-swept NMR spectra were taken by Fourier
transforming the spin-echo signal with the step-sum technique.
The value of $T_1^{-1}$ was determined by 
fitting the time dependence of the spin-echo intensity of the central transition line after 
the inversion pulse to the theoretical formula\cite{NarathTiNMR}. 
Good fitting was obtained in the whole temperature range, 4.2--300~K.

\begin{figure}[htb]
\centering
\includegraphics[width=1.0\linewidth]{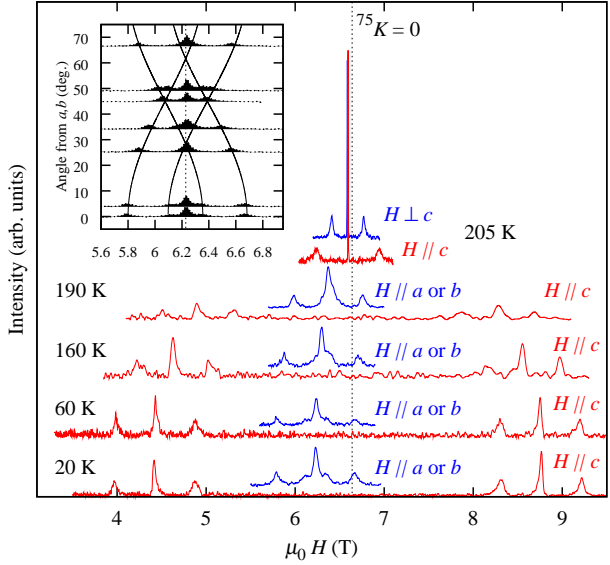}
\caption{(Color online) $^{75}$As-NMR spectra of SrFe$_2$As$_2$ obtained by
sweeping the magnetic field at the fixed frequency of 48.31~MHz along the
$c$-axis (red) or along the $a$- or $b$- axis (blue).
Below $T_\text{t}$, the staggered AF fields split the lines symmetrically
against the paramagnetic central position.
The inset shows angular variation of the NMR spectra at 20~K for the field
rotated in the $ab$-plane. Two sets of satellite lines appear due to twinning
in the orthorhombic structure.
The satellite positions are fitted to Eq.~\eqref{eq:res} with the quadrupole
parameters: $\nu^c = 3.31$~MHz, $|\nu^a - \nu^b|/|\nu^c| = 1.34$ (solid lines).}
\label{fig:spectra}
\end{figure}

%%\section{Experimental Results}
Figure~\ref{fig:spectra} shows the $^{75}$As-NMR spectra obtained by sweeping
the magnetic field. Since $^{75}$As nuclei have spin $3/2$, the NMR spectrum
consists of three transition lines. The central line appears at the magnetic
Zeeman frequency $\mu_0 \gamma_\text{N}H_\text{eff}$ and the two satellite lines
at $\mu_0 \gamma_\text{N}H_\text{eff} \pm \delta\nu$ split by the quadrupole
interaction. Here $\mu_0$ is the vacuum permeability, $\gamma_\text{N}/2\pi
= 7.29019$~MHz/T is the nuclear gyromagnetic ratio and $\bm H_\text{eff} = \bm H + \bm H_\text{hf}$ is sum of
the external field and the magnetic hyperfine field from neighboring Fe spins.
In the paramagnetic state, $H_\text{hf}$ is proportional to $H$, $H_\text{hf} =
K H$, $K$ being the Knight shift. The quarupole splitting follows the angular
dependance,
\begin{equation}\label{eq:res}
\delta\nu = \frac{1}{2}\left\{\nu^c(3\cos^2\theta - 1) + (\nu^a -
\nu^b)\sin^2\theta \cos 2\phi\right\},
\end{equation}
where $\theta$ is the angle between $\bm H_\text{eff}$ and the $c$-axis, $\phi$
is the azimuthal angle of $\bm H_\text{eff}$ in the $ab$-plane, and $\nu^\alpha
= eV_{\alpha\alpha}Q/2h$ with $e$, $V_{\alpha\alpha}$, $Q$, and $h$ being the
elementary charge, the electric field gradient (EFG), the nuclear quadrupole
moment, and the Planck's constant, respectively.

In the paramagnetic state $T > T_\text{t}$\,(=199~K), a
very sharp central line with the full width at the half maximum of
4.5~kHz is observed (Fig.~\ref{fig:spectra}).
Below $T_\text{t}$, the spectrum split into the two sets of three lines for
$H \parallel c$.
For $H \perp c$, in turn, the spectrum does not split but shifts to lower
fields.
This indicates that the AF order produces a staggered hyperfine field along the
$c$-axis, ${\bm H}_\text{hf} = (0, 0, \pm \Delta)$. 
Similar spectral changes were observed in BaFe$_2$As$_2$ and
we showed that only the stripe-type AF order can lead to such a hyperfine
field\cite{KitagawaBa122}.

\begin{figure}[tb]
\centering
\includegraphics[width=0.9\linewidth]{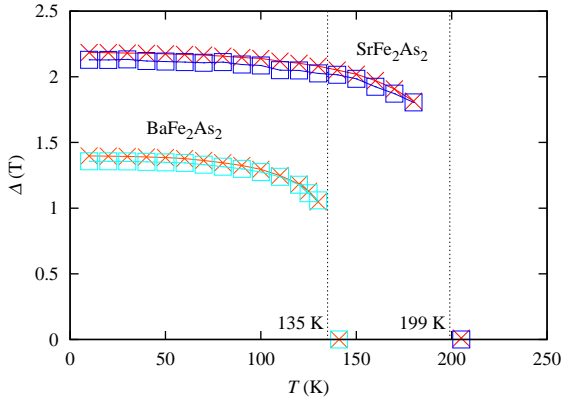}
\caption{(Color online) The temperature dependence of the hyperfine field
$\Delta$ at the As nuclei associated with the AF order.
$\Delta$ is determined by the splitting of the
central line for $H \parallel
c$ (crosshairs) or by the relation $H_\text{eff} = \sqrt{H^2
+ \Delta^2}$ for $H \perp c$ (squares)\cite{KitagawaBa122}.}
\label{fig:gap}
\end{figure}
The hyperfine field is generally expressed as the sum of contributions from
neighboring Fe spins $\bm S_{(i)}$, $\bm H_\text{hf}=\sum_i B^{(i)} \cdot g\bm
S_{(i)}$, where $B^{(i)}$ is the hyperfine coupling tensor to the spin at $i$-th
site, and $g$ is the $g$ factor. The expression can be converted to sum of the
Fourier components, $\bm H_\text{hf} = \sum_{\bm q} \bm H_\text{hf}(\bm q), \bm H_\text{hf}(\bm q)
= B(\bm q) \cdot g \bm S(\bm q)$.
For the stripe-type AF structure $\bm Q=(10l)$ (referred to the orthorhombic
reciprocal lattice), symmetry consideration concludes that only the
$ac$-component of $\bm B(\bm Q)$ is non-zero\cite{KitagawaBa122},
\begin{equation}\label{eq:chifetoas}
\left( \begin{array} {c}
H_\text{hf}^a(\bm Q) \\
H_\text{hf}^b(\bm Q) \\
H_\text{hf}^c(\bm Q)
\end{array} \right)
 =  
4g\left( \begin{array} {ccc}
0& 0& B_{ac} \\
0& 0& 0  \\
B_{ac}& 0& 0
\end{array} \right)
 \left( \begin{array} {c}
S^{a}(\bm Q) \\
S^{b}(\bm Q) \\
S^{c}(\bm Q) 
\end{array} \right).
\end{equation}
Then, $\Delta = 4B_{ac}\sigma_{a}$, where
$\sigma_{a}=g|S^{a}(\bm Q)|$ is the AF moment per site along
the $a$-direction in the unit of the Bohr's magneton $\mu_\text{B}$. The
temperature dependence of $\Delta$ is plotted in Fig.~\ref{fig:gap}. The jump of $\Delta$ at $T_\text{t}$ is
expected for the first-order transition.
From the ordered moments determined by the
neutron scattering measurements ($\sigma_{a}$=1.01~$\mu_\text{B}$ for
SrFe$_2$As$_2$\cite{KanekoSr122NS}, and 0.87~$\mu_\text{B}$ for
BaFe$_2$As$_2$\cite{HuangBa122NS}), $B_{ac}$ are determined as
0.53~T/$\mu_\text{B}$ for SrFe$_2$As$_2$ and
0.43~T/$\mu_\text{B}$ for BaFe$_2$As$_2$.
\begin{figure}[tb]
\centering
\includegraphics[width=0.9\linewidth]{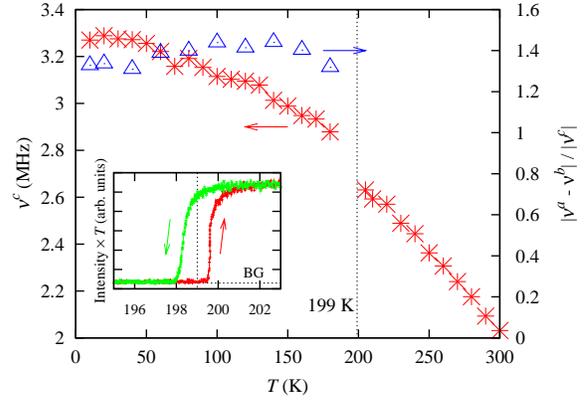}
\caption{(Color online) The quadrupole splitting frequency along the $c$
axis, $\nu^c$, and the asymmetry parameter $|\nu^a - \nu^b| / |\nu^c|$ are plotted as a function of temperature.  
The asymmetry parameter is zero in the tetragonal phase.
%The jump of
%$\nu^c$ and the emergence of the asymmetry indicate the first-order structural
%transition into the orthorhombic phase.
In the inset, the temperature
variation of the intensity of the central line for the paramagnetic phase is
shown for $H \perp c$. The vanishing intensity is due to the shift of the line
caused by the AF order as shown in Fig.~\ref{fig:spectra}. The abrupt change
with hysteresis is the evidence for the first-order magnetic transition.}
\label{fig:nuq}
\end{figure}

Figure~\ref{fig:nuq} shows the quadrupole splitting $\nu^{i}$ as a function of
temperature. In the tetragonal phase, the $c$-axis corresponds to the
largest principal value of the EFG,
and $2\nu^{a} = 2\nu^{b} = -\nu^{c}$.
Thus the asymmetric parameter $| \nu^a - \nu^b |/| \nu^c |$ is zero.
Below $T_\text{t}$, it shows a jump,
a direct evidence for the first-order structural transition. The value
of $| \nu^a - \nu^b |/| \nu^c |$ exceeding unity means that the principal axis
of the largest EFG rotates from the $c$-axis above $T_\text{t}$ to the $a$- or
$b$-axis below $T_\text{t}$. Such a drastic change of EFG has been also
observed in BaFe$_2$As$_2$. In the inset of Fig.~\ref{fig:nuq}, the peak intensity of the central line for $H
\perp c$ in the paramagnetic phase is plotted near the
transition. The hysteresis of about 1~K with the transitional width within 0.5~K
demonstrates good homogeneity of the sample.

\begin{figure}[tb]
\centering
\includegraphics[width=0.9\linewidth]{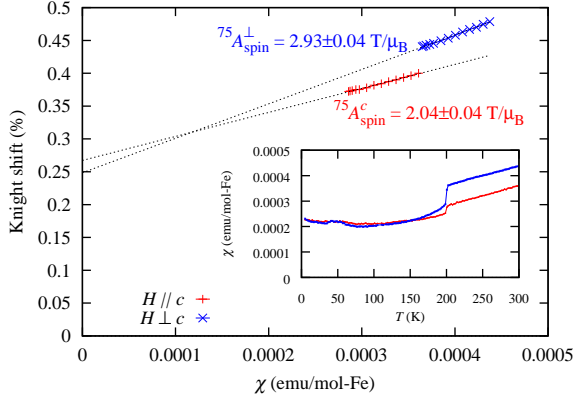}
\caption{(Color online) The Knight shift $^{75}K$ is plotted against the bulk
susceptibility $\chi$. The dotted lines represent the fits to a linear relation.
The inset shows the $\chi$ measured by a commercial SQUID magnetometer at
5~T.}
\label{fig:kchi}
\end{figure}
In Fig.~\ref{fig:kchi}, the Knight shift $^{75}K$ is plotted against the
susceptibility in the paramagnetic phase, after correcting for the
demagnetization field and the second order quadrupolar shift. In general, the Knight shift consists of the $T$-dependent 
spin shift, and the $T$-independent chemical (orbital)
shift, $K(T)=K_\mathrm{chem} + K_\mathrm{spin}(T)$.
The spin part of Knight shift is linearly related to the temperature-dependent
spin susceptibility $\chi_\mathrm{spin}(T)$ via the hyperfine coupling
tensor $B = \sum_i B_{(i)}$, $K^\alpha_\mathrm{spin}(T) = 
B_{\alpha\alpha}\chi^\alpha_\mathrm{spin}(T)/N_\text{A}\mu_\text{B}\,(\alpha=a,b,
\text{or}\,c)$. Here $N_\text{A}$ is the Abogadro's number. From the slope of
the plot in Fig.~\ref{fig:kchi}, the hyperfine coupling are obtained as $2.93\pm 0.04$~T/$\mu_\text{B}$ for $H \perp c$, and $2.04\pm
0.04$~T/$\mu_\text{B}$ for $H \parallel c$. 
The isotropic part $B_\text{iso} = (2B_{aa} + B_{cc})/3$, which originates
from the Fermi contact interaction with the As-$s$ orbitals, is three times
larger than the anisotropic part $B_\text{aniso} = B_{aa} - B_{cc}$.
The latter is mainly due to As-$p$ orbitals, which contribute to the conduction
bands through hybridization with the Fe-$d$ orbitals. Similar result was
also reported for BaFe$_2$As$_2$\cite{KitagawaBa122}.
Although the off-diagonal component $B_{ac}$ also originates from the $p$
orbitals, it cannot be determined from the present $K$-$\chi$ analysis, because
$B_{ac}$ does not contribute to the uniform ($\bm q= \bm 0$) hyperfine field.
It plays a key role, however, in the nuclear relaxation as we discuss below.

\begin{figure}[tb]
\centering
\includegraphics[width=1.0\linewidth]{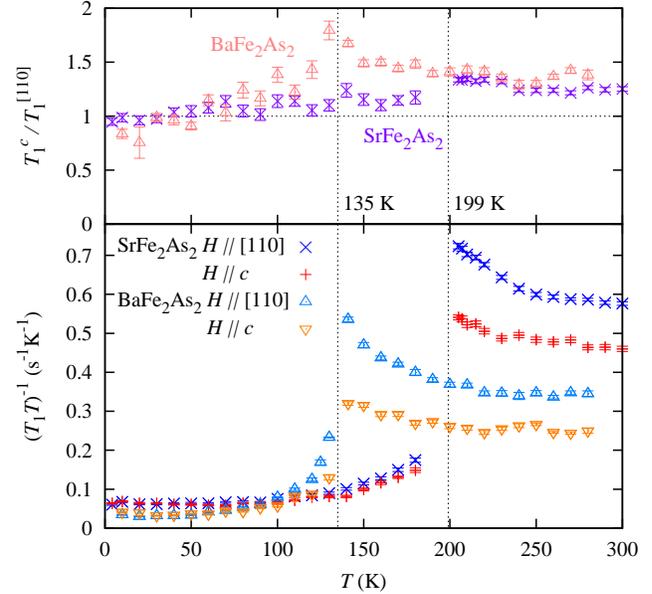}
\caption{(Color online) Lower panel: nuclear spin-lattice relaxation rate
divided by temperature, $(T_1T)^{-1}$, is plotted as a function of temperature for two field-orientations.
Orthorhombic notation is used for both above and below $T_\text{t}$ to keep
consistency. Upper panel: the temperature dependence of the anisotropy of
$(T_1T)^{-1}\,(=T_1^c / T_1^{[110]})$.
The results for BaFe$_2$As$_2$\cite{KitagawaBa122} are shown for comparisons.}
% is plotted, which is a good index for
%the dominance of the anisotropic stripe-type antiferromagnetic fluctuations
%with the wave vector of $(10l)$.}
\label{fig:t1t}
\end{figure}
Next, the magnetic fluctuations are discussed based on the results of the
spin-lattice relaxation rate $T_1^{-1}$.
%%Relaxation rate divided by
%%temperature $(T_1T)^{-1}$ is generally proportional to low-energy dynamic
%%susceptibility averaged in $q$ space with so-called form factors.
The lower panel of Fig.~\ref{fig:t1t} shows the temperature dependence of the
relaxation rate divided by temperature $(T_1T)^{-1}$ for SrFe$_2$As$_2$ and
BaFe$_2$As$_2$.
For both compounds, a clear reduction of $(T_1T)^{-1}$ is observed across the
transition.
At the lowest temperatures, $(T_1T)^{-1}$ becomes constant, which is
a specific feature of Fermi liquids. This indicates that small Fermi surfaces
remain in the AF state, which is consistent with the
quantum oscillation experiments\cite{AnalytisBa122QOSC,SebastianSr122QOSC}.
%an itinerant antiferromagnetism driven by Fermi-surface nesting.

The most prominent feature is the large enhancement of $(T_1T)^{-1}$ as the
temperature approaches $T_\text{t}$, in particular for $H \parallel c$. This
indicates development of strong AF fluctuations in the paramagnetic state, even
through critical slowing down of magnetic fluctuations is generally not
expected for a first-order phase transition. In order to see the
anisotropic behavior, we plotted the ratio $T_1^c/T_1^\text{[110]}$ against
temperature in the upper panel of Fig.~\ref{fig:t1t}. This shows that the ratio
increases significantly near $T_\text{t}$, i.\,e. the upturn of $(T_1T)^{-1}$
is anisotropic, in BaFe$_2$As$_2$. On the other hand, the ratio is nearly
independent of temperature in SrFe$_2$As$_2$. The different behavior for the
two materials can be understood qualitatively in terms of anisotropic AF
fluctuations as follows.

The nuclear relaxation rate can be expressed in terms of the fluctuations of
the hyperfine field perpendicular to the magnetic field at the NMR angular
frequency $\omega_\text{res}$. Since both the hyperfine coupling and the spin
correlation function are anisotropic, it is necessary to consider not only the imaginary
part of the Fe spin susceptibility $\text{Im}\chi^\perp(\omega_\text{res})$ but
the hyperfine field at the As site $H_\text{hf}^\perp(\omega_\text{res})$. Then,
\begin{align}
(T_1^z)^{-1}
 &= \frac{(\mu_0\,\gamma_\text{N})^2}{2}
\int^\infty_{-\infty} \text{d}t\, e^{i\omega_\text{res}
t}\big(
\langle\{H_\text{hf}^x(t), H_\text{hf}^x(0)\}\rangle\nonumber\\
&\quad\quad+\langle\{H_\text{hf}^y(t), H_\text{hf}^y(0)\}\rangle\big)\\
&={(\mu_0\,\gamma_\text{N})^2} \left\{
\left|H_\text{hf}^x(\omega_\text{res})\right|^2 +
\left|H_\text{hf}^y(\omega_\text{res})\right|^2 \right\}\\
\label{eq:t1}&={(\mu_0\,\gamma_\text{N})^2} \sum_{\bm q}\left\{
\left|H_\text{hf}^x(\bm q, \omega_\text{res})\right|^2 +
\left|H_\text{hf}^y(\bm q, \omega_\text{res})\right|^2 \right\},
\end{align}
when $z$ is the direction of the field and $\left|X(\omega)\right|^2$ denotes the power spectral
density of a time-dependent random variable $X(t)$. The enhancement of
$(T_1T)^{-1}$ should be ascribed to the spin fluctuations near the ordering
wave vector $\bm Q= (10l)$\cite{notet1}.
In fact, short range AF order at the same vector has been reported by
quasi-elastic neutron scattering in BaFe$_2$As$_2$\cite{MatanBa122INS}. We use
the orthorhombic notation both above and below $T_\text{t}$ to keep consistency.

If we consider only $\bm q = \bm Q$ contribution for simplicity, combining with
Eqs.~\eqref{eq:t1} and \eqref{eq:chifetoas}, the anisotropy of $T_1^{-1}$ is
given by
\begin{equation}
\left( \begin{array} {c}
(T_1^a)^{-1} \\
(T_1^b)^{-1} \\
(T_1^c)^{-1}
\end{array} \right)
 \propto  \left( \begin{array} {c}
\left|S^{a}(\bm Q, \omega_\text{res})\right|^2\\
\left|S^{a}(\bm Q, \omega_\text{res})\right|^2
 + \left|S^{c}(\bm Q, \omega_\text{res})\right|^2 \\
 \left|S^{c}(\bm Q, \omega_\text{res})\right|^2 \end{array} \right) .
\end{equation}
Here $S^i(\bm q,\omega)$ is the dynamical representation of $S^i(\bm q)$.
Note that the behavior $(T_1T)^{-1} \sim const.$ at high temperatures is due 
to contribution from a broad region in the $q$-space away from $\bm Q$, 
which is not included in the above expression.  Hnece the analysis is valid only 
qualitatively. 

In the paramagnetic tetragonal phase or for the case of $H \parallel [110]$ in
the orthorhombic phase, the in-plane anisotropy is averaged. Then,
\begin{equation}\label{eq:t1aniso}
\frac{(T_1^{\text{[110]}})^{-1}}{ (T_1^{c})^{-1} } = 
\frac{
2\left|S^{a}(\bm Q, \omega_\text{res})\right|^2
 + \left|S^{c}(\bm Q, \omega_\text{res})\right|^2 }
{2\left|S^{c}(\bm Q, \omega_\text{res})\right|^2}.
\end{equation}
The neutron diffraction experiments showed the ordered moments in the AF states
are directed along the $a$-axis\cite{KanekoSr122NS,HuangBa122NS}. If the spin
fluctuations above $T_\text{t}$ are strongly anisotropic [$|S^{a}(\omega)| \gg
|S^{c}(\omega)|$],
\begin{equation}
\frac{(^{75}T_1^{\text{[110]}})^{-1}}{ (^{75}T_1^{c})^{-1} } \gg 1.
\end{equation}
On the other hand, if the fluctuations are isotropic [$|S^{a}(\omega)| =
|S^{c}(\omega)|$],
\begin{equation}\label{eq:isotropic}
\frac{(T_1^{\text{[110]}})^{-1}}{ (T_1^{c})^{-1} } = \frac{3}{2}.
\end{equation}

The data in the upper panel of Fig.~\ref{fig:t1t} indicate that SrFe$_2$As$_2$
corresponds to the latter isotropic case, while in BaFe$_2$As$_2$ the ratio of
$T_1^{-1}$ exceeds 1.5 near $T_\text{t}$ suggesting more anisotropic AF
fluctuations. Such difference in the anisotropy of spin fluctuations indicates 
different roles of the spin-orbit interaction in the two materials. Specifically,  
the anisotropy of the spin fluctuations can be related to the orbital character
of the electronic states as follows. In iron pnictides, the five-fold
degeneracy of the $d$ orbitals are partially lifted near the Fermi level, the
$d_{xz}$, $d_{yz}$, and $d_{x^{2}-y^{2}}$ having the dominant
weight\cite{IkedaFLEX122}.
Generally the spin excitations are associated with either the intraband or
interband transition, which generates orbital fluctuations along specific directions.
For example, the transiton between $d_{xz}$ and $d_{yz}$ induces 
orbital fluctiaions along the $z$ direction, while the transition between
$d_{x^{2}-y^{2}}$ and $d_{xz}$ (or $d_{yz}$) generates fluctuations of $L_{x}$
and $L_{y}$.
The preferred direction of the orbital fluctiations thus determined by the
geometry and orbital characters of the Fermi surfaces will cause anisotropic
spin fluctuations via the spin-orbit interaction.
In fact, more detailed and quantitative analysis should be possible by using
the tight-binding representation of the band structure. Such an analysis is highly desired.  

%%\section{Conclusions}
In summary, we have investigated the detailed $^{75}$As-NMR studies in
SrFe$_2$As$_2$ at ambient pressure.
Clear evidence for the first-order structural and AF phase transition is
observed from the change of NMR spectra. The enhanced nuclear relaxation rate 
in the vicinity of the transition is most probably caused by the stripe AF
fluctuations. The anisotropy of $T_1^{-1}$ indicates that the stripe AF
fluctuations become anisotropic in the spin-space in BaFe$_2$As$_2$ near the
transition, but remains isotropic in SrFe$_2$As$_2$.
It is interesting to see how the AF fluctuations and their anisotropy
change when the materials become superconducting by pressure or by doping.

%%\section*{Acknowledgments}
We thank M.~Yoshida for helpful discussions. This work was
supported partly by the Grant-in-Aids on Priority Areas ``Invention of Anomalous Quantum Materials'' (No. 16076204),
by the Global COE program, and by Special Coordination Funds for Promoting
Science and Technology ``Promotion of Environmental Improvement for Independence of Young Researchers'' from MEXT of Japan. 
K.\,K. is financially supported as a JSPS research fellow.

\bibliography{document}

%%%%%%%%Bibiography Style File for JPSJ %%%%%%%%%%%%%
% Released on November 15, 1996: Version 1.00       %
% Copyright (C) 1996 by Shinsaku Fujita,            %
%                             all rights reserved.  %
%%%%%%%%%%Bibliography%%%%%%%%%%%%%%%%%%%%%%%%%%%%%%%
\begin{thebibliography}{10}

\bibitem{KamiharaJACS}
Y. Kamihara, T. Watanabe, M. Hirano, and H. Hosono:  J. Am. Chem. Soc. {\bf
  130} (2008) 3296.
\bibitem{RotterBa122SDW}
M. Rotter, M. Tegel, D. Johrendt, I. Schellenberg, W. Hermes, and R.
  P{\"o}ttgen:  Phys. Rev. B {\bf 78} (2008) 020503(R).
\bibitem{RotterBK122}
M. Rotter, M. Tegel, and D. Johrendt:  Phys. Rev. Lett. {\bf 101} (2008)
  107006.
\bibitem{SasmalSrK122}
K. Sasmal, B. Lv, B. Lorenz, A. Guloy, F. Chen, Y. Xue, and C.~W. Chu:  Phys.
  Rev. Lett. {\bf 101} (2008) 107007.
\bibitem{WuCaNa122}
G. Wu, H. Chen, T. Wu, Y.~L. Xie, Y.~J. Yan, R.~H. Liu, X.~F. Wang, J.~J. Ying,
  and X.~H. Chen:  J. Phys.:Condens. Matter {\bf 20} (2008) 422201.
\bibitem{KotegawaSr122HP}
H. Kotagawa, H. Sugawara, and H. Tou:  J. Phys. Soc. Jpn. {\bf 78} (2009)
  013709.
\bibitem{Alireza122HP}
P.~L. Alireza, Y.~T.~C. Ko, J. Gillett, C.~M. Petrone, J.~M. Cole, G.~G.
  Lonzarich, and S.~E. Sebastian:  J. Phys.: Condens. Matter {\bf 21} (2008)
  012208.
\bibitem{YuCa122HeHP}
W. Yu, A.~A. Aczel, T.~J. Williams, S.~L. Bud’ko, N. Ni, P.~C. Caneld, and
  G.~M. Luke:  cond-mat/0811.2554.
\bibitem{FukazawaBa122HP}
H. Fukazawa, N. Takeshita, T. Yamazaki, K. Kondo, K. Hirayama, Y. Kohori, K.
  Miyazawa, H. Kito, H. Eisaki, and A. Iyo:  J. Phys. Soc. Jpn. {\bf 77} (2008)
  105004.
\bibitem{HuangBa122NS}
Q. Huang, Y. Qiu, W. Bao, J. Lynn, M. Green, Y. Gasparovic, T. Wu, G. Wu, and
  X.~H. Chen:  Phys. Rev. Lett. {\bf 101} (2008) 257003.
\bibitem{KanekoSr122NS}
K. Kaneko, A. Hoser, N. Caroca-Canales, A. Jesche, C. Krellner, O. Stockert,
  and C. Geibel:  Phys. Rev. B {\bf 78} (2008) 212502.
\bibitem{KitagawaBa122}
K. Kitagawa, N. Katayama, K. Ohgushi, M. Yoshida, and M. Takigawa:  J. Phys.
  Soc. Jpn. {\bf 77} (2008) 114709.
\bibitem{YanSr122SC}
J.-Q. Yan, A. Kreyssig, S. Nandi, N. Ni, S.~L. Bud’ko, A. Kracher, R.~J.
  McQueeney, R.~W. McCallum, T.~A. Lograsso, A.~I. Goldman, and P.~C. Caneld:
  Phys. Rev. B {\bf 78} (2008) 024516.
\bibitem{NarathTiNMR}
A. Narath:  Phys. Rev. {\bf 162} (1967) 320.
\bibitem{AnalytisBa122QOSC}
J.~G. Analytis, R.~D. McDonald, J.-H. Chu, S.~C. Riggs, A.~F. Bangura, C.
  Kucharczyk, M. Johannes, and I.~R. Fisher:  cond-mat/0902.1172.
\bibitem{SebastianSr122QOSC}
S.~E. Sebastian, J. Gillett, N. Harrison, P.~H.~C. Lau, D.~J. Singh, C.~H.
  Mielke, and G.~G. Lonzarich:  J. Phys.: Condens. Matter {\bf 20} (2008)
  422203.
\bibitem{notet1}
The enhanced anisotropy near $T_\text{t}$ for BaFe$_2$As$_2$,
  $(T_1^{\perp})^{-1}/(T_1^{c})^{-1} \gg 1$ indicates
  $|H_\text{hf}^{c}(\omega)| \gg |H_\text{hf}^{\perp}(\omega)|$. Our argument
  in Ref.~\citen{KitagawaBa122} leads to two possibilities; two-dimensional
  ferromagnetic fluctuations, which is rather unlikely, or anisotropic stripe
  AF fluctuations.
\bibitem{MatanBa122INS}
K. Matan, R. Morinaga, K. Iida, and T.~J. Sato:  Phys. Rev. B {\bf 79} (2009)
  054526.
\bibitem{IkedaFLEX122}
H. Ikeda:  J. Phys. Soc. Jpn. {\bf 77} (2008) 123707.
\end{thebibliography}

\end{document}